\documentclass[conference]{IEEEtran}
\pdfoutput=1
\usepackage[hyphens]{url}
\usepackage{hyperref}
\hypersetup{breaklinks=true, colorlinks,allcolors=blue}
\usepackage[backend=biber, style=ieee]{biblatex}

\IEEEoverridecommandlockouts
\usepackage{amsmath,amssymb,amsfonts}
\usepackage{algorithmic}
\usepackage{graphicx}
\usepackage{textcomp}
\usepackage{xcolor}  
\usepackage{balance}
\usepackage{mathtools}  
\usepackage{orcidlink}  
\usepackage{kantlipsum} 

\def\BibTeX{{\rm B\kern-.05em{\sc i\kern-.025em b}\kern-.08em
    T\kern-.1667em\lower.7ex\hbox{E}\kern-.125emX}}

\begin{document}

\title{Toward an Insider Threat Education Platform: \\ A Theoretical Literature Review}

\newcommand{\comments}[1]{}

\author{

Haywood Gelman\,\orcidlink{0009-0009-7208-1624}\,\textsuperscript{†}, 
John D. Hastings\,\orcidlink{0000-0003-0871-3622}\,\textsuperscript{‡}
David Kenley\,\orcidlink{0000-0001-7780-7644}\,\textsuperscript{‡}, 
and Eleanor Loiacono\,\orcidlink{0000-0002-4059-702X}\,\textsuperscript{§}

  \thanks{\textsuperscript{†}The Beacom College of Computer \& Cyber Sciences, Dakota State University, Madison, SD, USA. Email: haywood.gelman@trojans.dsu.edu}
\thanks{\textsuperscript{‡}The Beacom College of Computer \& Cyber Sciences, Dakota State University, Madison, SD, USA. Email: john.hastings@dsu.edu}
\thanks{\textsuperscript{‡}College of Arts \& Sciences, Dakota State University, Madison, SD, USA. Email: david.kenley@dsu.edu}
\thanks{\textsuperscript{§}Mason School of Business, The College of William and Mary, Williamsburg, VA, USA. E-mail: eleanor.loiacono@mason.wm.edu}
}

\comments {
\author{\IEEEauthorblockN{1\textsuperscript{st} Given Name Surname}
\IEEEauthorblockA{\textit{dept. name of organization (of Aff.)} \\
\textit{name of organization (of Aff.)}\\
City, Country \\
email address or ORCID}
\and
\IEEEauthorblockN{2\textsuperscript{nd} Given Name Surname}
\IEEEauthorblockA{\textit{dept. name of organization (of Aff.)} \\
\textit{name of organization (of Aff.)}\\
City, Country \\
email address or ORCID}
}
}

\maketitle
\begin{abstract}
Insider threats (InTs) within organizations are small in number but have a disproportionate ability to damage systems, information, and infrastructure. Existing InT research studies the problem from psychological, technical, and educational perspectives. Proposed theories include research on psychological indicators, machine learning, user behavioral log analysis, and educational methods to teach employees recognition and mitigation techniques. Because InTs are a human problem, training methods that address InT detection from a behavioral perspective are critical.  While numerous technological and psychological theories exist on detection, prevention, and mitigation, few training methods prioritize psychological indicators. This literature review studied peer-reviewed, InT research organized by subtopic and extracted critical theories from psychological, technical, and educational disciplines. In doing so, this is the first study to comprehensively organize research across all three approaches in a manner which properly informs the development of an InT education platform. 
\end{abstract}


\section{Introduction}

Insider threats (InTs) are among the most impactful risks facing organizations today. Insiders present a great threat to organizations due to their sanctioned access to systems, information, and infrastructure and the damage that can occur should they decide to act against the interests of an employer~\cite{greitzer_identifying_2012}. Recent statistics highlight the severity of the issue: in 2023, InTs were responsible for approximately 20\% of attacks on organizations across a range of attack vectors, while 74\% of attacks involved human interaction~\cite{verizon_dbir_dbir_2023}. This suggests that insider threats are not only technical but also deeply human problems, influenced by a range of psychological, behavioral, and organizational factors.

The increasing prevalence of InT incidents, rising from 68\% in 2021 to 74\% in 2023~\cite{cybersecurity_insiders_2023_2023}, underscores the urgent need for organizations to implement robust detection and mitigation strategies. These strategies must encompass both technological solutions and human-centered approaches to detect and respond to potential threats effectively. However, while significant advances have been made in technological detection systems, such as machine learning and user behavior analytics, the psychological perspective of insider threats remains underexplored~\cite{cappelli_cert_2012}. This study aims to address this gap by classifying theories related to developing the necessary training methods to effectively detect InTs with a particular emphasis on the psychological perspectives. In doing so, this is the first study to comprehensively organize research related to psychological, technological, and educational approaches.

\section{Methodology}

This study conducted a theoretical literature review that assessed psychological, technical, and educational perspectives on InT theories and subtopics. \cite{machi_literature_2022,kumar_research_2019} guided the approach to safeguard validity, minimize theoretical misinterpretation through triangulation, and to avoid selection bias through applicable search term construction. Content for theory extraction was sourced from searches conducted in peer-reviewed journals and conference papers using IEEE Xplore, ACM, Web of Science, and Google Scholar. Keywords included “psych”, “educ”, ontology”, “framework”, “model”, “persona”, “PMT”, “trust”, “machine”, “big data”, “behav”, “analytic”, “user analytic” combined with “inside” and “insider threat”. The progressive nature of the field negated the need for a publication date range, which enabled inclusion of three seminal psychological texts~\cite{bloom_taxonomy_1956,cressey_other_1953,mccrae_introduction_1992}. Theoretical extraction by application yielded common subtopics across psychology, technology, and education domains for analysis. The following sections detail the findings organized by psychological, technological, and educational approaches.

\section{Psychological InT Theories}

Psychological theories for InTs address characteristics, motivations, behaviors (CMB)~\cite{mills_current_2018}, and dark traits~\cite{maasberg_dark_2015} used to identify organizational threats. \cite{mundie_toward_2013} distinguished insiders from InTs based on an employee’s sanctioned access to systems, information, and infrastructure, their legitimate physical or virtual presence within the organization’s management boundaries, and InT potential to cause organizational harm. \cite{cappelli_management_2007} set precedent for~\cite{mundie_toward_2013} in stating the juxtaposition of InT attacks are in proximity to adverse employment incidents and work dissolution.

\subsection{InT Ontology}
InT ontology described an InT lexicon to define crucial aspects of InT terminology. Unintentional InT was similarly determined by \cite{schoenherr_multiple_2022} and \cite{greitzer_analysis_2014} to include carelessness, lack of training, and susceptibility to social engineering. \cite{yeo_unintentional_2023}, citing \cite{cappelli_cert_2012}, added that unintentional InT actions could be caused through ``action or inaction without malicious intent.''

Intentional threats are distinguished from unintentional threats by their mens rea \cite{schoenherr_multiple_2022}. \cite{yang_potential_2018}, citing \cite{mccrae_introduction_1992} and \cite{paulhus_dark_2002}, described intentional InT by the Five Factor Model and dark traits. The Five Factor Model included ``extraversion, agreeableness, conscientiousness, neuroticism, and openness to experience'' \cite{mccrae_introduction_1992}.

Dark traits were described by \cite{paulhus_dark_2002} as sinister aspects of personality considered to be below a level required for diagnosis but can be observed symptomatically \cite{paulhus_dark_2002}.  Dark traits include narcissism (self-interest), Machiavellianism (scheming), and psychopathy (lack of self-control) \cite{paulhus_dark_2002,harms_exposing_2022}. \cite{harms_exposing_2022} noted that due to the broad range of identification in the Five Factor Model, the model had not found wide acceptance in InT research circles, validated by \cite{maasberg_dark_2015}.  \cite{maasberg_dark_2015} also noted the inclusion of empathy and sense of entitlement to InT characteristics outside of dark traits \cite{maasberg_dark_2015}.

\cite{padayachee_conceptual_2013} introduced the concept of  InT triggered by opportunity to commit an InT action. \cite{padayachee_conceptual_2013} also cited \cite{cornish_opportunities_2003}, stating that “both motivation and opportunity play a role in crime; however, opportunity may be the ‘trigger’ to committing a crime.” \cite{padayachee_conceptual_2013}. \cite{padayachee_joint_2021} also described similarities to Cressey's Fraud Triangle \cite{cressey_other_1953}: behavioral characteristics that enable an InT to self-incentivize to commit an act, possessing the foresight to perceive an opening, and  the ability to justify the behavior.

\subsection{Frameworks and Models}
InT frameworks and models are methods for creating predictability in InT detections. \cite{nurse_understanding_2014} proposed the implementation of a detection model based on a combination of CMB of attackers, simulated attacks, and case studies. Research focused on building a library of case studies, dissecting them for their characteristic value, and extracting predictability features.

Where \cite{nurse_understanding_2014} concentrated on malicious and accidental InT attacks, \cite{padayachee_conceptual_2013} addressed threats from an opportunity landscape perspective. Citing \cite{cornish_opportunities_2003}, \cite{padayachee_conceptual_2013} noted the presence of incentive and prospectus in the commission of a crime, connecting these factors to InT who are catalysts to commit an InT act but wait for the prospect to surface. \cite{padayachee_framework_2015} modified this theory, granting heavier weight to prospectus over incentive. Padayachee’s position was contradicted by \cite{apau_theoretical_2019}, stating three well-known factors of crime commission – means, motive, and opportunity – where the reduction of one reduced the likelihood and/or impact of an InT attack.

Rather than addressing the criminological perspective of opportunity-based threats, \cite{ikany_symptomatic_2019} proposed a framework for InT detection based on behavioral indicators. Technology and non-technology based indicators were compiled from case studies to create an ontology of InT to inform the model. The research used the risk management model OCTAVE to validate indicators, although the model developers caution~\cite{caralli_introducing_2007} that it is not specified for InT risk management.

In contrast to \cite{ikany_symptomatic_2019} and \cite{padayachee_conceptual_2013,padayachee_framework_2015},  \cite{eom_framework_2011} proposed a conventional layered defense model through technological and user entity behavioral anomaly detection.

\subsection{Personas}
In addition to opportunity-based threats, unintentional and intentional threats address the most commonly seen InTs in organizations. \cite{cert_insider_threat_team_unintentional_2013} \cite{maasberg_dark_2015} CERT Insider Threat Team described CMB of unintentional InTs to include carelessness, lacking mindfulness of surroundings, and reduced awareness of consequences of unconscious compromise. \cite{greitzer_analysis_2014}  and  \cite{yeo_unintentional_2023} agreed with \cite{cert_insider_threat_team_unintentional_2013}, but also determined that unintentional InTs can cause damage by either active participation or inactivity.

\cite{schoenherr_cybersecurity_2021} and \cite{schoenherr_multiple_2022} described the Cybersecurity Questionnaire (CSEC) as an instrument to determine the level of risk an unintentional InT poses to an organization based on cyber habits. Contrary to \cite{greitzer_analysis_2014}, \cite{schoenherr_multiple_2022} noted the characteristic of communal involvement, where such activity could be viewed negatively in relation to social engineering.

\cite{uebelacker_social_2014} expanded on social engineering, explaining that the susceptibility of individuals to the act was related to six ideologies of influence: authority, commitment, reciprocity, liking, social proof, and scarcity.

Discussing intentional InTs, \cite{robayo_enemy_2022} described the decision to commit a malicious act as predicated on a perceived injustice, progressing through several stages before completion and flight. \cite{maasberg_dark_2015} identified the traits necessary to commit an intentional InT act as dark traits: non-criminal characteristics that enable an individual to bypass morality, willfully cause damage, and consciously plan an escape.

Dark traits were extended by \cite{harms_exposing_2022} to include a complex set of associated emotions and behaviors, applicable to both intentional and unintentional InTs.  \cite{padayachee_joint_2021} assented on emotional complexity, adding cognitive nullification as a negative counterbalance to committing an InT crime.

Protection Motivation Theory (PMT) applied to intentional InTs exhibited unique properties. According to \cite{vetter_assessing_2022}, PMT psychological behavior applied to those in a position to cause harm chose to protect individuals and organizations instead. This unique property demonstrates an opposing view to \cite{maasberg_dark_2015} and \cite{robayo_enemy_2022}. \cite{humaidi_procedural_2022} and \cite{posey_protection-motivated_2010} concurred.

\subsection{Prevention}
Prevention strategies seek to identify methods to mitigate InT attacks. \cite{alsowail_techniques_2022} organized InT defense into three categories, ``detection approaches, detection \& prevention approaches, or prevention approaches.'' 

\cite{greitzer_identifying_2012} identified a detection approach based on psychological antecedents. Their method assessed emotional state, complaints outside the workplace, distraction, and other characteristics that were indicative of an impending InT attack.

\cite{herath_encouraging_2009} proposed a prevention strategy based on stress from organizational peer relationships and consequences enforced by the organization for violations. The research concluded that InT behaviors can be manipulated by internal and external factors.

\cite{huertas-baker_exploring_2022} agreed with \cite{herath_encouraging_2009} on influencing factors, adding psychological triggers as a preceding act. \cite{johnston_fear_2010} studied the preventive impact of  distress petitions on the protection response of organizational IT communities, finding a positive correlation.

\cite{sperry_confronting_2014} proposed a detection and prevention approach by identifying numerous work and personal indicators of stress and compromise. Although \cite{sperry_confronting_2014}'s indicators were accurately identified, research minimally described noted traits as psychologically impacting, departing from the focus of the paper. \cite{voss_insider_2023} appraised performance management, control failure, behavioral moderation, authentication policy, and preemptive detection.

\section{Technological Theories}
\comments {
  \textcolor{blue}{Integrate:} \textbf{Technological frameworks employ machine learning models described by Padmavathi et al. \cite{padmavathi_framework_2022}, Reinerman-Jones et al. \cite{reinerman-jones_scenarios_2017}, and Yousef et al. \cite{yousef_machine_2023}. They also describe system logging by He et al. \cite{he_insider_2021}, and Khaliq et al. \cite{khaliq_role_2020}, authentication tracking by Nithiyanandam et al. \cite{nithiyanandam_advanced_2012}, and Tupakula \& Varadharajan \cite{tupakula_trust_2013}, while relying on computer-aided detection.}
  }

Technological theories are models that utilize technology to aid in InT discovery and mitigation. Subtopics include data-driven and machine-learning, trusted user approaches, and user behavior analytics. 

\subsection{Data-Driven \& Machine Learning Theories}
\cite{jabbour_insider_2010} proposed a system-level framework that detected and prevented unauthorized database changes through an automated security-defense mechanism based on administrative policy.  \cite{basu_towards_2018} presented an approach directly applicable to detection through behavioral analysis and game theory simulation. \cite{basu_towards_2018} implemented machine learning for data analysis but used game simulation to collect data for the model.

\cite{brdiczka_proactive_2012} also proposed a game theory approach using a World of Warcraft dataset to preemptively model InT behavior through a machine learning approach applied to psychological profiling and graph learning. Using structural anomaly detection (analysis of normal computer-use behavior) and psychological profiling (assessment of anomalous behavior arising out of computer use tracking), the model effectively predicted InT behavior when combining indicators.

\cite{ma_dante_2020} proposed creation of a neural network, DANTE, for detecting InT anomalous behavior in system logs.  Although DANTE showed a 93\% success rate detecting InTs, the dataset was simulated \cite{cert_int_dataset_insider_2020}, and it was only successful in identifying known attacks \cite{ma_dante_2020}.

\cite{martinez-moyano_modeling_2006} employed a system dynamics model to assess vulnerabilities in enterprise environments.  Through the researchers application of their dynamic trigger hypothesis (assessment of inferential indicators that create an environment in which InTs can thrive) the model was simulated to stimulate the surfacing of psychological precursors to vulnerabilities. Research concluded that technology and behavioral modeling of InTs were only as effective as humans ability to interpret them.

\cite{reinerman-jones_scenarios_2017} proposed the use of a machine learning simulation for their InT immersion study but with participant observational awareness and a requirement to act with fraudulent intent during the study. Simulation demonstrated a positive result for InT detection.

\cite{mayhew_use_2015} concurred on the use of machine learning to overcome the non-real-time nature of IT vulnerability processing. \cite{parveen_unsupervised_2012} observed higher correlation for InT detection with unsupervised learning over supervised machine learning models.

\subsection{Trust Theories}
Trust theories involve human-dependent access levels to systems, information, and infrastructure relative to human reliability \cite{colwill_human_2009}. Organizational trust levels increase as reliability increases, while simultaneously demonstrating an exponential increase in the relationship between trust levels and damage caused by trusted insiders \cite{colwill_human_2009}.

\cite{paci_detecting_2013} proposed use of the SI* modeling language to address InT behavior using an asset, role, and behavior-based framework to establish simulated trust levels between objects in a patient observation setting. The model proposed a design to enable IT staff to model user behavior on systems and permissions to preemptively determine individuals as malicious insiders.

\cite{sood_exploiting_2017} described a taxonomy for exploiting trusted social networking applications and the insertion of social networking malware (socioware) to steal personal information.  In contrast to~\cite{paci_detecting_2013} and~\cite{sood_exploiting_2017}, \cite{aldairi_trust_2019} proposed an unsupervised machine learning model to mine system logs for untrusted activities, noting \cite{parveen_insider_2011} observed higher correlation with unsupervised learning.

\cite{apau_theoretical_2019} studied a trust-matrix framework based on characteristics derived from CERT’s Common Sense Guide to Mitigating Insider Threats dataset~\cite{cert_int_dataset_insider_2020}. The model distilled trust aspects, character and competence, into behaviors associated with intent, integrity, capability, and results across a behavioral lifespan. The study instrument demonstrated positive correlation for the model.

\cite{ho_behavioral_2009} assessed trustworthiness through the perception of humans-as-sensors, trained to observe anomalous behavior, akin to a firewall. The research determined a positive correlation between trained human observers and detection of InTs.

\cite{li_towards_2020} discovered a positive correlation between distributed intrusion detection appliances and InT detection when implementing multi-level trust across collective detectors.  In their qualitative study and in contrast to technological models,~\cite{rousseau_insider_2021} addressed the replacement of the trusted computing model to reduce the risk of InTs through participant interviews.

\subsection{User Behavioral Analytics Theories}
User behavioral analytics theories model and analyze human behavior through unsupervised machine learning, user roles, behavioral profiling \& historical analysis, and user intent. \cite{khaliq_role_2020}'s literature review described industry approaches to user entity behavioral analysis (UEBA) tools that use supervised and unsupervised machine learning to mine logging facilities for evidence of InT behavior, but did not provide a novel approach to solve the problem.

Similar to \cite{apau_theoretical_2019} and \cite{ma_dante_2020}, \cite{he_insider_2021} used CERT’s Common Sense Guide to Mitigating Insider Threats dataset \cite{cert_int_dataset_insider_2020} to model InT behavior. Unlike \cite{apau_theoretical_2019}'s assessment of trust behaviors and \cite{ma_dante_2020}'s known-behavior limitation, \cite{he_insider_2021} trained a deep-learning model on historical user behavioral sequencing to achieve a 99.15\% detection success rate.

\cite{reinerman-jones_scenarios_2017} approach to environmental simulation enlisted participants with awareness of the experiment but not the scenario to encourage expected InT behaviors. Participants were required to commit actions designed to trigger machine learning recognition.

\cite{singh_user_2021} designed a model based on a human-driven augmented decision making framework.  In their model, data extraction methods used by \cite{he_insider_2021}, in conjunction with the CERT InT dataset \cite{cert_int_dataset_insider_2020}, and an unsupervised deep-learning model similar to \cite{aldairi_trust_2019}, and \cite{parveen_unsupervised_2012}, produced improved results over existing test methods at a mean of 84.12\%.

\section{Educational Frameworks and Models}

Educational frameworks and models refer to theories applied to teaching security education, training, and awareness (SETA) concepts to an array of participant profiles. To assess educational frameworks and models for InTs, it is necessary to include generalized SETA so a complete set of theories can be drawn. \cite{ops_executive_2011} and \cite{ops_presidential_2012} reformed national cybersecurity protection initiatives specific to InTs. Also, \cite{walden_national_2023} provides a strong foundation for generalizable SETA that can be modified for InT education purposes. The NICE Framework~\cite{petersen_workforce_2020} presents a competency and role-based training solution founded on task, knowledge, and skill objectives. In addition, CDSE Insider Threat training~\cite{cdse_int_training_insider_2023}, the MERIT program~\cite{cappelli_management_2007} and the CERT Guide to Insider Threats~\cite{cappelli_cert_2012}  are crucial examples of effective InT training. These generalized and InT-specific education initiatives embody critical perspectives that directly address this research's focus. The following section enumerates important educational frameworks and models applicable to psychology-focused InT training.

\cite{ramsoonder_optimizing_2020} applied Bloom’s Taxonomy of cognition \cite{bloom_taxonomy_1956} to cyber education to assess learning levels and outcomes mapped to NICE Framework objectives. \cite{burns_assessing_2015} applied expectancy theory to SETA to influence “security-related behavior”. Expectancy theory involves inducing an outcome by influencing the factors that create it, similar to PMT \cite{posey_protection-motivated_2010,vetter_assessing_2022} but with a control. \cite{gundu_ignorance_2013} concurred on PMT, while including Theory of Reasoned Action and Behaviorism Theory. \cite{burns_assessing_2015} discovered positive correlation between SETA and insider security behaviors.

\cite{cappelli_management_2007} presented a SETA guide on InTs specific to sabotage in the workplace. The model utilized psychological aspects of InTs, along with CMB, in a design to create predictive and preventive behaviors in an organization. ~\cite{hobbs_insider_2015} employed a case study approach to train nuclear power system educators on how to teach their nuclear system operator classes in the United Kingdom.

When teaching in a heterogeneous environment, it is important to employ training methods that accommodate different knowledge absorption techniques, according to the Felder-Silverman learning style model (FSLSM) \cite{graf_analysing_2007}. FSLSM can also be seen in the interactive, adventure-driven video game CyberCIEGE~\cite{cone_video_2007}. The game was used to successfully teach SETA to Navy students through scenarios and problem solving exercises \cite{cone_video_2007}. \cite{kwon_enriching_2017} added hands-on lab exercises to further improve SETA’s cognitive impact.

\section{Summary of InT Theories}

This study extracted, compared, and contrasted psychological, technical, and educational InT theories for future synthesis with an InT psychological-focused training initiatives. Supporting psychological theories include CMB, dark traits and cognitive nullification, organizational harm, adverse employment incidents, unintentional InTs and social engineering, intentional InTs, opportunity-based InTs, PMT, prevention categories, antecedents, peer relationships and stress, distress petitions, and preemptive action.

Technological theories include InT behavior modeling, system log behavioral analysis, observational awareness, unsupervised machine learning, trusted insiders, trusted behaviors, historical user behavior deep-learning, and trained human behaviors.

Educational theories include generalized and InT SETA training, Bloom’s taxonomy of cognition, security-related behavior, case study analysis, adjusting SETA methods to learners knowledge absorption methods, and hands-on lab exercises.

\section{Conclusion}
A large body of knowledge exists on the psychology of InTs. Research addressed existing studies, frameworks, ontologies, and theories that propose identification of InTs from characteristics, motivations, behaviors, and dark traits of InT personas. An equally large body of knowledge exists on technological means of detecting InTs through behavioral and observational analysis, machine learning, and trusted behaviors. Organizations rely heavily upon technology to spot indicators of InTs, most notably when addressing an entity that can be inherently unpredictable. Numerous guidelines exist on how organizations should train employees on InTs, but gaps in training methods possess an imbalance between psychological and technical detection and mitigation methods. Future research will create an awareness training platform for organizations on InTs based on psychology first and technology second. The conceptual framework will draw upon determiners of InT activity from psychological, technological, and educational theories extracted in this paper and implemented as part of a comprehensive InT training platform. 

\printbibliography

@techreport{petersen_workforce_2020,
	title = {Workforce Framework for Cybersecurity ({NICE} Framework)},
	%url = {https://nvlpubs.nist.gov/nistpubs/SpecialPublications/NIST.SP.800-181r1.pdf},
	abstract = {This publication from the National Initiative for Cybersecurity Education (NICE) describes the Workforce Framework for Cybersecurity (NICE Framework), a fundamental reference for describing and sharing information about cybersecurity work. It expresses that work as Task statements and describes Knowledge and Skill statements that provide a foundation for learners including students, job seekers, and employees. The use of these statements helps students to develop skills, job seekers to demonstrate competencies, and employees to accomplish tasks. As a common, consistent lexicon that categorizes and describes cybersecurity work, the NICE Framework improves communication about how to identify, recruit, develop, and retain cybersecurity talent. The NICE Framework is a reference source from which organizations or sectors can develop additional publications or tools that meet their needs to define or provide guidance on different aspects of cybersecurity education, training, and workforce development.},
	%language = {en},
	%urldate = {2022-09-26},
	%institution = {National Institute of Standards and Technology},
	institution = {{NIST}},
	author = {Petersen, Rodney and Santos, Danielle and Smith, Matthew C. and Wetzel, Karen A. and Witte, Greg},
	%month = nov,
	year = {2020},
	doi = {10.6028/NIST.SP.800-181r1},
	file = {Petersen et al. - 2020 - Workforce Framework for Cybersecurity (NICE Framew.pdf:/Users/hgelman/Zotero/storage/4C7FAHGC/Petersen et al. - 2020 - Workforce Framework for Cybersecurity (NICE Framew.pdf:application/pdf},
}

@mastersthesis{mills_current_2018,

@article{schoenherr_multiple_2022,
	title = {Multiple Approach Paths to Insider Threat ({MAP}-{IT}): Intentional, Ambivalent and Unintentional Insider Threats},
	volume = {1},
	shorttitle = {Multiple {Approach} {Paths} to {Insider} {Threat} ({MAP}-{IT})},
	url = {https://citrap.scholasticahq.com/article/37117},
	abstract = {Insider threats (InT) are a growing concern for private and public institutions, resulting in a shift of emphasis from perimeter-based defences to internal detection mechanisms. Many approaches that address InT assume that these are pathological behaviors, perpetrated by misanthropic ‘malicious insiders’. We present a novel interdisciplinary framework (Multiple Approach Paths to Insider Threat, or MAP-IT) that emphasizes the importance of both individual motivation and social context. Rather than assuming InTs reflect a homogenous ill-intentioned group of individuals that deviate from organizational norms, we consider the importance of general social psychological and personality factors for detecting and responding to InT, especially within the Western intelligence and security context. MAP-IT is based on the premise that InTs can be separated into three motivational pathways (intentional, unintentional, or ambivalent) and that the intentional pathway can be further subdivided into prosocial, asocial, and antisocial motivation.},
	%language = {en},
	number = {1},
	urldate = {2022-10-06},
	journal = {Counter-Insider Threat Research and Practice},
	author = {Schoenherr, Jordan Richard and Lilja-Lolax, Kristoffer and Gioe, David},
	month = aug,
	year = {2022},
	file = {Full Text PDF:/Users/hgelman/Zotero/storage/8CU5JYTU/Schoenherr et al. - 2022 - Multiple Approach Paths to Insider Threat (MAP-IT).pdf:application/pdf;Snapshot:/Users/hgelman/Zotero/storage/D735YXKE/37117.html:text/html},
}

@misc{ops_presidential_2012,
	title = {Presidential {Memorandum} -- {National} {Insider} {Threat} {Policy} and {Minimum} {Standards} for {Executive} {Branch} {Insider} {Threat} {Programs}},
	url = {https://obamawhitehouse.archives.gov/the-press-office/2012/11/21/presidential-memorandum-national-insider-threat-policy-and-minimum-stand},
	abstract = {MEMORANDUM FOR THE HEADS OF EXECUTIVE DEPARTMENTS AND AGENCIES SUBJECT: National Insider Threat Policy and Minimum Standards for Executive Branch Insider Threat Programs},
	%language = {en},
	urldate = {2022-10-20},
	journal = {whitehouse.gov},
	author = {OPS},
	month = nov,
	year = {2012},
	file = {Snapshot:/Users/hgelman/Zotero/storage/7VWHFVFI/presidential-memorandum-national-insider-threat-policy-and-minimum-stand.html:text/html},
}

@article{harms_exposing_2022,
	title = {Exposing the darkness within: {A} review of dark personality traits, models, and measures and their relationship to insider threats},
	volume = {71},
	issn = {22142126},
	shorttitle = {Exposing the darkness within},
	%url = {https://linkinghub.elsevier.com/retrieve/pii/S2214212622002228},
	doi = {10.1016/j.jisa.2022.103378},
	abstract = {Insider threats are a pernicious threat to modern organizations that involve individuals intentionally or unin­ tentionally engaging in behaviors that undermine or abuse information security. Previous research has estab­ lished that personality factors are an important determinant of the likelihood that an individual will engage in insider threat behaviors. The present article asserts that dark personality traits, non-clinical personality char­ acteristics that are typically associated with patterns of anti-social and otherwise noxious interpersonal behav­ iors, may be particularly useful for understanding and predicting insider threat behaviors. Although some relationships between insider threats and dark traits have been documented, most attention has been devoted to a limited subset of dark traits. To address this issue, we critically review contemporary models of dark traits and their potential value for understanding both malicious and non-malicious insider threats, supplemented by discussions of subject matter expert ratings concerning the relevance of dark traits for both insider threat be­ haviors and cybersecurity personnel job performance. We then review potential assessment issues and provide evidence of possible moderators for the relationships under investigation. Finally, we develop avenues for future research, an agenda for improving the measurement of dark traits, and guidance for how organizations may implement the assessment of dark traits in their organizational processes.},
	%language = {en},
	%urldate = {2023-06-01},
	journal = {Journal of Information Security and Applications},
	author = {Harms, P.D. and Marbut, Alexander and Johnston, Allen C. and Lester, Paul and Fezzey, Tyler},
	%month = dec,
	year = {2022},
	pages = {103378},
	file = {Harms et al. - 2022 - Exposing the darkness within A review of dark per.pdf:/Users/hgelman/Zotero/storage/YDZP9KAZ/Harms et al. - 2022 - Exposing the darkness within A review of dark per.pdf:application/pdf},
}

@article{paulhus_dark_2002,
	title = {The {Dark} {Triad} of personality: {Narcissism}, {Machiavellianism}, and psychopathy},
	volume = {36},
	issn = {00926566},
	shorttitle = {The {Dark} {Triad} of personality},
	%url = {https://linkinghub.elsevier.com/retrieve/pii/S0092656602005056},
	doi = {10.1016/S0092-6566(02)00505-6},
	abstract = {Of the oﬀensive yet non-pathological personalities in the literature, three are especially prominent: Machiavellianism, subclinical narcissism, and subclinical psychopathy. We evaluated the recent contention that, in normal samples, this ÔDark TriadÕ of constructs are one and the same. In a sample of 245 students, we measured the three constructs with standard measures and examined a variety of laboratory and self-report correlates. The measures were moderately inter-correlated, but certainly were not equivalent. Their only common Big Five correlate was disagreeableness. Subclinical psychopaths were distinguished by low neuroticism; Machiavellians, and psychopaths were low in conscientiousness; narcissism showed small positive associations with cognitive ability. Narcissists and, to a lesser extent, psychopaths exhibited self-enhancement on two objectively scored indexes. We conclude that the Dark Triad of personalities, as currently measured, are overlapping but distinct constructs.},
	%language = {en},
	number = {6},
	%urldate = {2023-06-11},
	journal = {Journal of Research in Personality},
	author = {Paulhus, Delroy L and Williams, Kevin M},
	%month = dec,
	year = {2002},
	pages = {556--563},
	file = {Paulhus and Williams - 2002 - The Dark Triad of personality Narcissism, Machiav.pdf:/Users/hgelman/Zotero/storage/6Q3YCTM4/Paulhus and Williams - 2002 - The Dark Triad of personality Narcissism, Machiav.pdf:application/pdf},
}

@report{verizon_dbir_dbir_2023,
	title = {2023 Data Breach Investigations Report ({DBIR})},
	url = {https://www.verizon.com/business/resources/T31a/reports/2023-data-breach-investigations-report-dbir.pdf},
	author = {{Verizon}},
	year = {2023},
	%file = {2023-data-breach-investigations-report-dbir.pdf:/Users/hgelman/Zotero/storage/L76LAQPW/2023-data-breach-investigations-report-dbir.pdf:application/pdf},
    urldate={2024-10-12}
}

@inproceedings{mundie_toward_2013,
	%address = {New Orleans, LA, USA},
	title = {Toward an {Ontology} for {Insider} {Threat} {Research}: {Varieties} of {Insider} {Threat} {Definitions}},
	%isbn = {978-0-7695-5065-7},
	shorttitle = {Toward an {Ontology} for {Insider} {Threat} {Research}},
	%url = {https://ieeexplore.ieee.org/document/6691366},
	doi = {10.1109/STAST.2013.14},
	abstract = {The lack of standardization of the terms insider and insider threat has been a noted problem for researchers in the insider threat field. This paper describes the investigation of 42 different definitions of the terms insider and insider threat, with the goal of better understanding the current conceptual model of insider threat and facilitating communication in the research community.},
	%language = {en},
	%urldate = {2023-09-23},
	booktitle = {2013 {Third} {Workshop} on {Socio}-{Technical} {Aspects} in {Security} and {Trust}},
	publisher = {IEEE},
	author = {Mundie, David A. and Perl, Sam and Huth, Carly L.},
	%month = jun,
	year = {2013},
	pages = {26--36},
	file = {Mundie et al. - 2013 - Toward an Ontology for Insider Threat Research Va.pdf:/Users/hgelman/Zotero/storage/NL79GTAD/Mundie et al. - 2013 - Toward an Ontology for Insider Threat Research Va.pdf:application/pdf},
}

@inproceedings{burns_assessing_2015,
	%address = {HI, USA},
	title = {Assessing the {Role} of {Security} {Education}, {Training}, and {Awareness} on {Insiders}' {Security}-{Related} {Behavior}: {An} {Expectancy} {Theory} {Approach}},
	%isbn = {978-1-4799-7367-5},
	shorttitle = {Assessing the {Role} of {Security} {Education}, {Training}, and {Awareness} on {Insiders}' {Security}-{Related} {Behavior}},
	%url = {http://ieeexplore.ieee.org/document/7070290/},
	doi = {10.1109/HICSS.2015.471},
	abstract = {Organizational success in the digital age is largely dependent upon the ability to collect, manage, and transfer proprietary information. Given this knowledge economy, it is no exaggeration to say that the protection of sensitive information is a top priority for most firms. However, achieving information security is complicated by the increased access to organizationally relevant information afforded to employees—putting organizational information security largely at the mercy of insiders.},
	%language = {en},
	%urldate = {2023-09-23},
	%booktitle = {Proceedings of the 2015 48th {Hawaii} {International} {Conference} on {System} {Sciences}},
	booktitle = {48th Hawaii Int. Conf. on {System} {Sciences}},
	publisher = {IEEE},
	author = {Burns, A.J. and Roberts, Tom L. and Posey, Clay and Bennett, Rebecca J. and Courtney, James F.},
	%month = jan,
	year = {2015},
	%pages = {3930--3940},
	%file = {Burns et al. - 2015 - Assessing the Role of Security Education, Training.pdf:/Users/hgelman/Zotero/storage/95W696UX/Burns et al. - 2015 - Assessing the Role of Security Education, Training.pdf:application/pdf},
}

@inproceedings{greitzer_analysis_2014,
	%address = {San Jose, CA},
	title = {Analysis of Unintentional Insider Threats Deriving from Social Engineering Exploits},
	%isbn = {978-1-4799-5103-1},
	%url = {http://ieeexplore.ieee.org/document/6957309/},
	doi = {10.1109/SPW.2014.39},
	abstract = {Organizations often suffer harm from individuals who bear no malice against them but whose actions unintentionally expose the organizations to risk—the unintentional insider threat (UIT). In this paper we examine UIT cases that derive from social engineering exploits. We report on our efforts to collect and analyze data from UIT social engineering incidents to identify possible behavioral and technical patterns and to inform future research and development of UIT mitigation strategies.},
	%language = {en},
	%urldate = {2023-09-23},
	booktitle = {2014 {IEEE} {Security} and {Privacy} {Workshops}},
	publisher = {IEEE},
	author = {Greitzer, Frank L. and Strozer, Jeremy R. and Cohen, Sholom and Moore, Andrew P. and Mundie, David and Cowley, Jennifer},
	%month = may,
	year = {2014},
	pages = {236--250},
	%file = {Greitzer et al. - 2014 - Analysis of Unintentional Insider Threats Deriving.pdf:/Users/hgelman/Zotero/storage/47GYTH38/Greitzer et al. - 2014 - Analysis of Unintentional Insider Threats Deriving.pdf:application/pdf},
}

@inproceedings{yang_potential_2018,
	%address = {Bamberg},
	title = {Potential {Malicious} {Insiders} {Detection} {Based} on a {Comprehensive} {Security} {Psychological} {Model}},
	%isbn = {978-1-5386-5119-3},
	%url = {https://ieeexplore.ieee.org/document/8405686/},
	doi = {10.1109/BigDataService.2018.00011},
	abstract = {The insider threat continues to be a paramount cyber security challenge that threatens individuals, ﬁnancial enterprises and governmental organizations. To deter insider threats, traditional detection, which mainly focuses on policy checks and anomaly detection for users’ computers and network activities, has been studied widely. However, because insiders have intrinsic authorized access at attack under normal behavior proﬁles, it is necessary to integrate the attackers’ psychological characteristics. This work proposes a novel detection approach for potential malicious insiders based on a comprehensive security psychological model derived from Big-5 and Dark Triad personality traits, overcoming the biased choice and equality hypothesis problems in previous work. Moreover, the threat conﬁdence degree is proposed to identify pseudo abnormal users and to markedly reduce the false positive rate. The experimental results illustrate the effectiveness and feasibility of the proposed approach, which has a very low false negative rate, and lay the foundation for a promising insider threat detection approach that integrates the attackers’ psychological traits with the attack-chain characteristics.},
	%language = {en},
	%urldate = {2023-09-23},
	booktitle = {2018 {IEEE} {Fourth} {International} {Conference} on {Big} {Data} {Computing} {Service} and {Applications} ({BigDataService})},
	publisher = {IEEE},
	author = {Yang, Guang and Cai, Lijun and Yu, Aimin and Ma, Jiangang and Meng, Dan and Wu, Yu},
	%month = mar,
	year = {2018},
	pages = {9--16},
	file = {Yang et al. - 2018 - Potential Malicious Insiders Detection Based on a .pdf:/Users/hgelman/Zotero/storage/RNDHEF2X/Yang et al. - 2018 - Potential Malicious Insiders Detection Based on a .pdf:application/pdf},
}

@inproceedings{padayachee_joint_2021,
	address = {Durban, South Africa},
	title = {Joint {Effects} of {Neutralisation} {Techniques} and the {Dark} {Triad} of {Personality} {Traits} on {Gender} : {An} {Insider} {Threat} {Perspective}},
	%isbn = {978-1-72818-081-6},
	shorttitle = {Joint {Effects} of {Neutralisation} {Techniques} and the {Dark} {Triad} of {Personality} {Traits} on {Gender}},
	%url = {https://ieeexplore.ieee.org/document/9395053/},
	doi = {10.1109/ICTAS50802.2021.9395053},
	abstract = {Addressing the problem of the insider threat is multifaceted as it involves trusted individuals who have legitimate authorisation to the information resources of an organisation. It has been reasoned that understanding the catalysts, that is, motivation, rationalisation and opportunity that converge in a crime scenario, may assist in managing the problem. However, not all individuals who are motivated and in the presence of an opportunity for a crime will be maleficent. It is the rationalisation that intercedes in the decision-making step as to whether to commit a crime or not. Criminals must neutralise their feelings towards conformity using rationalisations or more specifically apply “techniques of neutralisation” (i.e. a set of rationalisation techniques), to engage in criminal activities. As the thought patterns and behaviour associated with the successful application of neutralisation techniques, appear to be characteristic of maladaptive behaviour, it may be contended that personality is a mediating factor in the process. This notion underpinned the theoretical basis of this research study which is founded on the socially aversive traits of the Dark Triad (DT), which may enable a dissociation with conformity thereby enhancing an insiders’ capacity to invoke techniques of neutralisation. However extant research shows these traits may be influenced by gender variations. Consequently, the goal of this research is to understand the joint effect of the traits of the DT and the techniques of neutralisation as moderated by gender from the perspective of an insider threat. The findings of this study may help in improving the management of insider threats.},
	%language = {en},
	%urldate = {2023-09-23},
	booktitle = {2021 {Conference} on {Information} {Communications} {Technology} and {Society} ({ICTAS})},
	publisher = {IEEE},
	author = {Padayachee, Keshnee},
	%month = mar,
	year = {2021},
	pages = {40--45},
	file = {Padayachee - 2021 - Joint Effects of Neutralisation Techniques and the.pdf:/Users/hgelman/Zotero/storage/98GIUTM6/Padayachee - 2021 - Joint Effects of Neutralisation Techniques and the.pdf:application/pdf},
}

@inproceedings{apau_theoretical_2019,
	%address = {Negeri Sembilan, Malaysia},
	title = {A {Theoretical} {Review}: {Risk} {Mitigation} {Through} {Trusted} {Human} {Framework} for {Insider} {Threats}},
	%isbn = {978-1-72815-657-6},
	shorttitle = {A {Theoretical} {Review}},
	%url = {https://ieeexplore.ieee.org/document/8970795/},
	doi = {10.1109/ICoCSec47621.2019.8970795},
	abstract = {This paper discusses the possible effort to mitigate insider threats risk and aim to inspire organizations to consider identifying insider threats as one of the risks in the company’s enterprise risk management activities. The paper suggests Trusted Human Framework (THF) as the on-going and cyclic process to detect and deter potential employees who bound to become the fraudster or perpetrator violating the access and trust given. The mitigation’s control statements were derived from the recommended practices in the “Common Sense Guide to Mitigating Insider Threats” produced by the Software Engineering Institute, Carnegie Mellon University (SEI-CMU). The statements validated via a survey which was responded by fifty respondents who work in Malaysia.},
	%language = {en},
	%urldate = {2023-09-23},
	booktitle = {2019 {International} {Conference} on {Cybersecurity} ({ICoCSec})},
	publisher = {IEEE},
	author = {Apau, Mohd Nazer and Sedek, Muliati and Ahmad, Rabiah},
	%month = sep,
	year = {2019},
	%pages = {37--42},
	%file = {Apau et al. - 2019 - A Theoretical Review Risk Mitigation Through Trus.pdf:/Users/hgelman/Zotero/storage/ULGT253I/Apau et al. - 2019 - A Theoretical Review Risk Mitigation Through Trus.pdf:application/pdf},
}

@inproceedings{aldairi_trust_2019,
	address = {Los Angeles, CA, USA},
	title = {A {Trust} {Aware} {Unsupervised} {Learning} {Approach} for {Insider} {Threat} {Detection}},
	%isbn = {978-1-72811-337-1},
	%url = {https://ieeexplore.ieee.org/document/8843465/},
	doi = {10.1109/IRI.2019.00027},
	abstract = {With the rapidly increasing connectivity in cyberspace, Insider Threat is becoming a huge concern. Insider threat detection from system logs poses a tremendous challenge for human analysts. Analyzing log ﬁles of an organization is a key component of an insider threat detection and mitigation program. Emerging machine learning approaches show tremendous potential for performing complex and challenging data analysis tasks that would beneﬁt the next generation of insider threat detection systems. However, with huge sets of heterogeneous data to analyze, applying machine learning techniques effectively and efﬁciently to such a complex problem is not straightforward.},
	%language = {en},
	%urldate = {2023-09-23},
	booktitle = {2019 {IEEE} 20th {International} {Conference} on {Information} {Reuse} and {Integration} for {Data} {Science} ({IRI})},
	publisher = {IEEE},
	author = {Aldairi, Maryam and Karimi, Leila and Joshi, James},
	%month = jul,
	year = {2019},
	pages = {89--98},
	file = {Aldairi et al. - 2019 - A Trust Aware Unsupervised Learning Approach for I.pdf:/Users/hgelman/Zotero/storage/G56Q87FR/Aldairi et al. - 2019 - A Trust Aware Unsupervised Learning Approach for I.pdf:application/pdf},
}

@inproceedings{tupakula_trust_2013,
	address = {Melbourne, Australia},
	title = {Trust {Enhanced} {Security} {Architecture} for {Detecting} {Insider} {Threats}},
	%isbn = {978-0-7695-5022-0},
	%url = {http://ieeexplore.ieee.org/document/6680886/},
	doi = {10.1109/TrustCom.2013.8},
	abstract = {Attacks on the organization networks can be classified as external and internal attacks. For the purpose of this paper we consider that external attacks are generated by the attackers or from hosts outside the organization, and internal attacks are generated by malicious insiders within the organization. Insider attacks have always been challenging to deal with as insiders have legitimate and physical access to the systems within the organization, and they have knowledge of the organization networks and more importantly, are aware of the security environment enforced within the organization. In this paper we propose novel trust enhanced security techniques to deal with the insider attack problem. Our architecture detects the attacks by monitoring the user activity as well as the state of the system using trusted computing in exposing and analyzing suspicious behaviour. We will demonstrate how an insider can exploit the weakness in the systems to generate different attacks and how our architecture can help to prevent such attacks.},
	%language = {en},
	%urldate = {2023-09-23},
	booktitle = {2013 12th {IEEE} {International} {Conference} on {Trust}, {Security} and {Privacy} in {Computing} and {Communications}},
	publisher = {IEEE},
	author = {Tupakula, Udaya and Varadharajan, Vijay},
	%month = jul,
	year = {2013},
	pages = {552--559},
	file = {Tupakula and Varadharajan - 2013 - Trust Enhanced Security Architecture for Detecting.pdf:/Users/hgelman/Zotero/storage/BFKNPRVF/Tupakula and Varadharajan - 2013 - Trust Enhanced Security Architecture for Detecting.pdf:application/pdf},
}

@inproceedings{maasberg_dark_2015,
	%address = {HI, USA},
	title = {The {Dark} {Side} of the {Insider}: {Detecting} the {Insider} {Threat} through {Examination} of {Dark} {Triad} {Personality} {Traits}},
	%isbn = {978-1-4799-7367-5},
	shorttitle = {The {Dark} {Side} of the {Insider}},
	%url = {http://ieeexplore.ieee.org/document/7070238/},
	doi = {10.1109/HICSS.2015.423},
	abstract = {Efforts to understand what goes on in the mind of an insider have taken a back seat to developing technical controls, yet insider threat incidents persist. We examine insider threat incidents with malicious intent and propose an explanation through a relationship between Dark Triad personality traits and the insider threat. Although Dark Triad personality traits have emerged in insider threat cases and deviant workplace behavior studies, they have not been labeled as such and little empirical research has examined this phenomenon. This paper builds on previous research on insider threat and introduces ten propositions concerning the relationship between Dark Triad personality traits and insider threat behavior. We include behavioral antecedents based on the Theory of Planned Behavior and Capability Means Opportunity (CMO) model and the factors affecting those antecedents. This research addresses the behavioral aspect of the insider threat and provides new information in support of academics and practitioners.},
	%language = {en},
	%urldate = {2023-09-23},
	booktitle = {Proceedings of the 2015 48th {Hawaii} {International} {Conference} on {System} {Sciences}},
	publisher = {IEEE},
	author = {Maasberg, Michele and Warren, John and Beebe, Nicole L.},
	%month = jan,
	year = {2015},
	pages = {3518--3526},
	file = {Maasberg et al. - 2015 - The Dark Side of the Insider Detecting the Inside.pdf:/Users/hgelman/Zotero/storage/4BCW2P3T/Maasberg et al. - 2015 - The Dark Side of the Insider Detecting the Inside.pdf:application/pdf},
}

@article{sood_exploiting_2017,
	title = {Exploiting {Trust}: {Stealthy} {Attacks} {Through} {Socioware} and {Insider} {Threats}},
	volume = {11},
	%issn = {1932-8184, 1937-9234, 2373-7816},
	shorttitle = {Exploiting {Trust}},
	%url = {http://ieeexplore.ieee.org/document/7042925/},
	doi = {10.1109/JSYST.2015.2388707},
	abstract = {Online social networks (OSNs) provide a new dimension to people’s lives by giving birth to online societies. OSNs have revolutionized the human experience, but they have also created a platform for attackers to distribute infections and conduct cybercrime. An OSN provides an opportunistic attack platform for cybercriminals through which they can spread infections at a large scale. We describe a category of malware (or attacks) known as socioware that exploits OSN environments for performing unauthorized and nefarious activities. Socioware can be an executable, an extension, an exploit code, etc., that conducts malicious operations in OSNs with serious impact on users. Furthermore, we discuss the socioware taxonomy highlighting the characteristics of socioware to illustrate the design and exploitation tactics of OSN malware. In contrast, insider threats (employees or contractors) are posing a grave threat to organizations, with a motivation to steal critical data and monetize it for ﬁnancial gains. Insider threats have become a serious concern for many organizations today. We present a complete attack model to demonstrate how an insider threat exploits the online trust and conﬁdentiality by transforming an OSN into a socioware distribution platform that infects other employees’ systems. Finally, we discuss security defenses that can be adopted to defend against socioware.},
	%language = {en},
	number = {2},
	%urldate = {2023-09-23},
	journal = {IEEE Systems Journal},
	author = {Sood, Aditya K. and Zeadally, Sherali and Bansal, Rohit},
	%month = jun,
	year = {2017},
	pages = {415--426},
	file = {Sood et al. - 2017 - Exploiting Trust Stealthy Attacks Through Sociowa.pdf:/Users/hgelman/Zotero/storage/CFB3XPDJ/Sood et al. - 2017 - Exploiting Trust Stealthy Attacks Through Sociowa.pdf:application/pdf},
}

@inproceedings{eom_framework_2011,
	address = {Piscataway, NJ},
	title = {A {Framework} of {Defense} {System} for {Prevention} of {Insider}'s {Malicious} {Behaviors}},
	isbn = {978-89-5519-155-4}, % 978-1-4244-8830-8}

@inproceedings{martinez-moyano_modeling_2006,
	address = {Monterey, CA, USA},
	title = {Modeling the {Emergence} of {Insider} {Threat} {Vulnerabilities}},
	%isbn = {978-1-4244-0501-5}, % 978-1-4244-0500-8}

@misc{hobbs_insider_2015,
	title = {Insider {Threats}: {An} {Educational} {Handbook} of {Nuclear} \& {Non}-{Nuclear} {Case} {Studies}},
	url = {https://www.kcl.ac.uk/csss/assets/insider-threats-handbook.pdf},
	publisher = {King's College London Centre for Science and Security Studies},
	author = {Hobbs, Christopher and Moran, Matthew},
	month = aug,
	year = {2015},
}

\end{document}